\documentstyle[epsf,aps,prb,multicol]{revtex}
\begin{document}
\newcommand{\Dvec}{{\rm\bf D}}
\newcommand{\Pvec}{{\rm\bf P}}
\newcommand{\Evec}{{\rm\bf E}}
\newcommand{\veps}{\varepsilon}  
\newcommand{\De}{$\Delta$}
\newcommand{\de}{$\delta$}
\newcommand{\mc}{\multicolumn}
\newcommand{\be}{\begin{eqnarray}}
\newcommand{\ee}{\end{eqnarray}}
\newcommand{\einf}{\varepsilon^\infty}
\newcommand{\ez}{\varepsilon^0}  
\newcommand{\strain}{{\bf \epsilon}}  
\newcommand{\eps}{\varepsilon}  
\renewcommand{\cite}[1]{[$\!\!$\onlinecite{#1}]}

\draft 
\title{SPONTANEOUS VS. PIEZOELECTRIC POLARIZATION 
IN III-V NITRIDES: CONCEPTUAL ASPECTS AND PRACTICAL CONSEQUENCES}  
\author{\underline{Fabio Bernardini}$^1$ and Vincenzo Fiorentini$^{1,2}$}
\address{$^1$Istituto Nazionale per la Fisica della Materia 
-- Dipartimento di 
 Fisica, Universit\`a di Cagliari, Cagliari, Italy;\\
Tel: +39-0706754847; Fax: +39-070570071; 
e-mail: fabio.bernardini@dsf.unica.it; \\
$^2$Walter Schottky Institut, Technische Universit\"a{}t
M\"u{}nchen, Garching, Germany.}

\maketitle

\begin{abstract}
Macroscopic polarization plays a major role in determining
the optical and electrical properties of nitride nanostructures
via polarization-induced built-in electrostatic fields. While
currently fashionable, this field of endeavour is still by far
 in its early infancy.
Here we contribute some clarifications on the conceptual issues
involved  in determining   built-in fields in III-V nitride
nanostructures, sorting out in particular the roles of spontaneous and 
piezoelectric polarization.
\end{abstract}



\begin{multicols}{2}
\section{INTRODUCTION}
 Macroscopic  polarization in   low-symmetry crystals
is well  known in the ferroelectrics community. 
Outside that community, it 
was so far held for a curiosity devoid of practical, 
let alone technological, relevance. 
In particular,
it was known  \cite{Catellani}
that a large {\it spontaneous} polarization (henceforth SP), 
at most an order of magnitude  smaller than
the giant values of ferroelectric perovskites, 
was present in some wurtzite semiconductors,
but due to their minor technological relevance 
 compared to zincblende III-V arsenides or phosphides, 
 this issue was largely overlooked outside a limited 
circle of specialists.

Such state of affairs has changed after first-principles 
calculations \cite{noi.pol} provided evidence of the 
 considerable SP and larger-then-usual piezoelectric 
coupling constants of {\it wurtzite III-V nitrides}.
Because of the huge  importance of the latter as materials 
for optoelectronic devices in the blue--UV range \cite{ambacher},
 the  issue of 
spontaneous polarization fields was  brought to a larger community.  

The obvious relevance of polarization in nitrides is due to the
fact that it induces large ($\sim$MV/cm)
built-in electrostatic fields  in layered nitride nanostructures,
as first shown in Ref. \cite{noi.int}. 
These fields affect dramatically the optical and electrical
properties of those structures [5-14]. Quite a flurry of activity 
has  recently revealed a host of new and unexpected effects:
to name but some,
Stark-like red shifts in recombination and absorption 
energies for increasing quantum well widths 
\cite{takeuchi,altri,langer,ganapl,noi.prb},
 concurrent suppression of oscillator strength \cite{altri,ganapl,noi.prb}, 
ensuing anomalies in recombination dynamics
\cite{altri,chichibu,gil,cingo}, and their   
interplay with charge injection
\cite{ganapl,noi.prb,nichia} driving a recovery of oscillator
strength and a blue back-shift of transition 
energies,
``self-doping'' effects at HEMT heterointerfaces \cite{oa},
have been demonstrated theoretically and experimentally.

In view of these exciting developments, it is likely that most
 of the potentially observable polarization-related effects 
in nitrides are still to be dreamt about. It is safe to state,
in any event, that conceptual
clarity on the determination of the electric fields produced
by polarization, and in particular by SP, is a must if we are to
understand and exploit these effects in practical applications.
To mention just a pair of issues, the built-in fields
 depend on boundary conditions and on device design (number of quantum
wells, thickness, composition, contacts, etc.);
 also, the field values and sign patterns are not those
expected if only piezoelectricity is considered.
This work therefore aims at clarifying some of these subtle 
questions, focusing on three inter-related systems, {\it i.e.}
large samples, multi-quantum-wells (MQWs),
and isolated quantum-wells (QWs).

\section{SPONTANEOUS VS. PIEZOELECTRIC}
\begin{figure}[h]
\epsfclipon
\epsfxsize=14cm
\vspace*{2cm}
\centerline{\epsffile{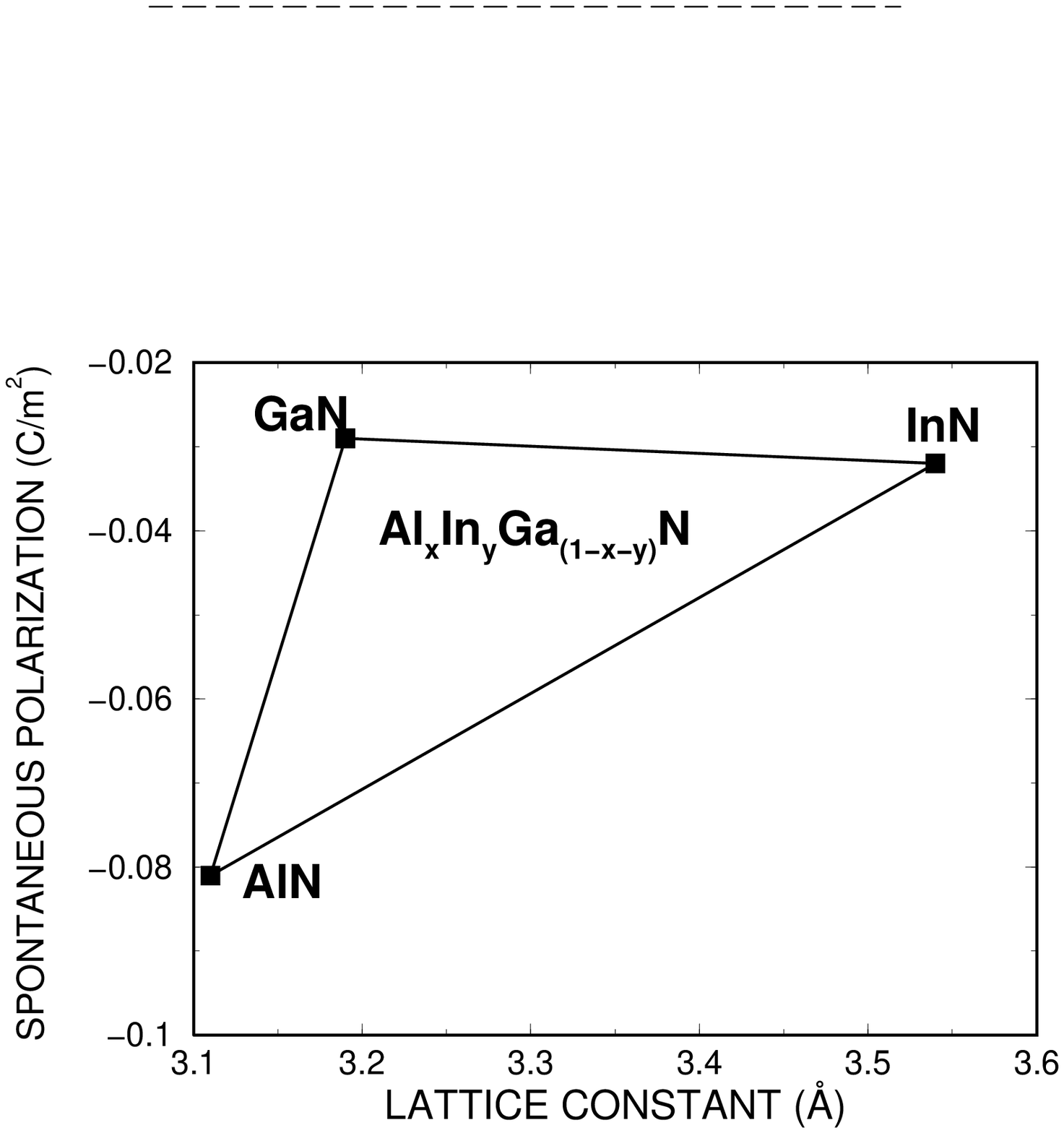}}
\vspace{3cm}
\caption{Spontaneous polarization in Al$_x$In$_y$Ga$_{1-x-y}$N alloys
according to a Vegard-like rule.}
\label{fig.spont}
\end{figure}

\begin{figure}[h]
\epsfclipon
\epsfxsize=14cm
\centerline{\epsffile{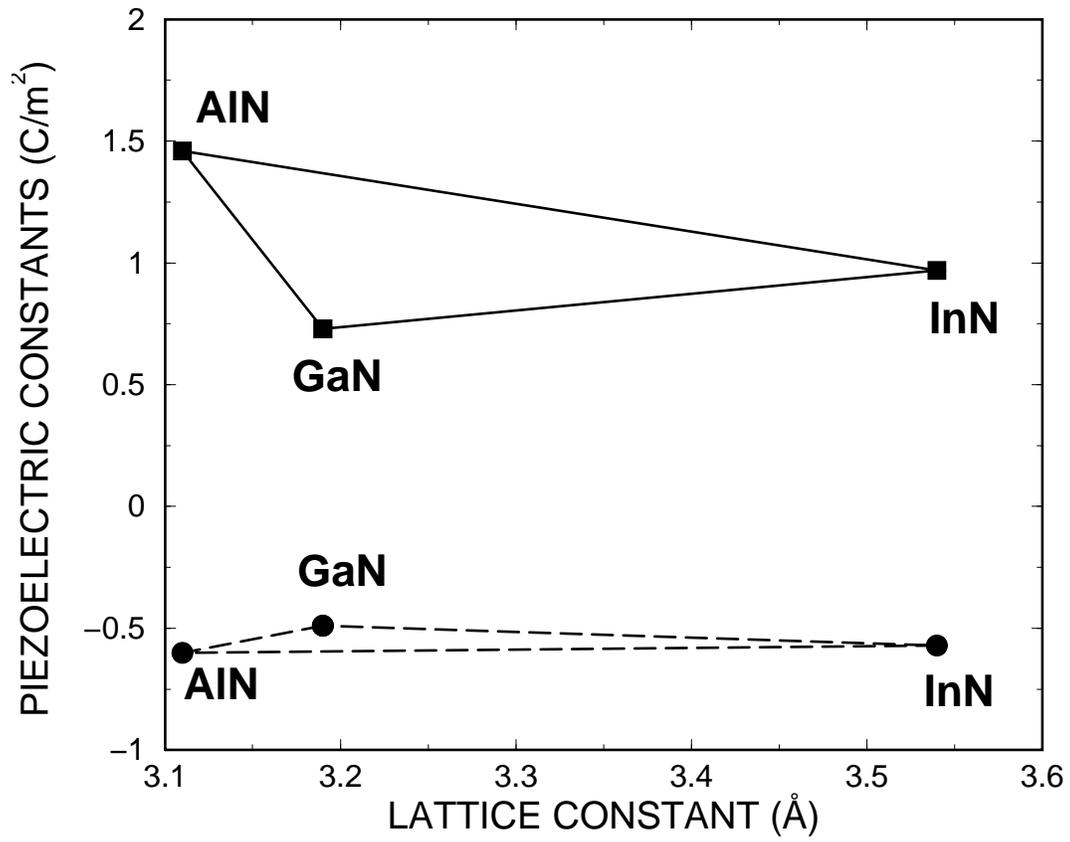}}
\vspace{3cm}
\caption{Piezoelectric coupling constants e$_{33}$ (squares)
and e$_{31}$ (dots)
after Ref.~\protect\onlinecite{noi.pol}.
Triangles refer to AlInGaN alloys
according to a Vegard-like rule.}
\label{fig.piezo}
\end{figure}

Spontaneous polarization 
and piezoelectric constants can be reliably determined 
 \cite{KS}
via simple  electronic structure calculations based on the
Berry  geometric quantum phase concept.
We start by discussing our recent result thereon \cite{noi.pol},
which are summarized in Figs. \ref{fig.spont} and  \ref{fig.piezo}.
The values  calculated ab initio are for III-N binaries (AlN, GaN,
InN) only; since direct calculations  for the 
alloys are not yet available, an 
estimate of the SP $\Pvec^{(\rm sp)}$ in Al$_x$In$_y$Ga$_{(1-x-y)}$N
alloys was obtained by  Vegard   interpolation:
\be
 \Pvec^{\rm (sp)}({\rm Al_xIn_yGa_{(1-x-y)}N}) &=&
 x ~\Pvec^{\rm (sp)}_{\rm AlN} + y ~\Pvec^{\rm (sp)}_{\rm InN} \\
                         &+& (1-x-y) ~\Pvec^{\rm (sp)}_{\rm GaN}\, ,
\nonumber
\ee
 similar relations holding for the piezoelectric constants.
Thus, the triangle in Fig.~\ref{fig.spont} borders the 
(theoretical) SP values achievable in a general III-N alloy 
 as a function of in-plane lattice constant. The triangles in
Fig. \ref{fig.piezo} border the accessible values of the two
piezoconstants relevant to $a$-plane epitaxial strains.  
The  large range of accessible polarizations and lattice constants is due to
the concurrent  large lattice mismatch between GaN and InN on one
side, and the very large SP in AlN on the other. 

Three main points are  evident from the data reported.
{\it First}, the piezoelectric constants are an order of 
magnitude larger than in other III-V's, and reversed in sign.   
{\it Second}, 
for  basal-plane strains  
typical of epitaxial wurtzite nitride multilayers
(2-5\%),  
SP is comparable to, or larger than piezoelectric
polarization. Therefore, SP is all but 
negligible in any typical III-V nitride nanostructures. 
{\it Third}, GaN has nearly the same SP as InN, but a large
lattice mismatch to it; on the other hand, AlN has a SP about 
three times larger than GaN, but a much smaller mismatch.
This implies that InGaN/GaN structures will tend to be mostly
influenced by piezoelectricity, whereas in
AlGaN/GaN systems, SP effects will be dominant. Mixed regimes can also
occur, which we will come back to in Sec.~\ref{sqw}.

\section{MASSIVE SAMPLES}
\label{massive}

By elementary electrostatics,  
polarization and electrostatic field are related by
\be
     \Evec = (\Dvec - \Pvec)/\eps
\label{field}
\ee
where $\Dvec$ is the displacement field, $\Pvec$ the macroscopic 
polarization, and $\eps$ the static dielectric constant.
The latter have been calculated \cite{noi.eps} to be
10.31, 10.28, and 14.61 for AlN, GaN, and InN respectively;
the value for generic alloys are estimated by Vegard 
interpolation.
The polarization in Eq. \ref{field} is  known as transverse,
 and  it is the sum of the spontaneous 
and piezoelectric polarization contributions: 
$\Pvec = \Pvec^{(\rm sp)} + \Pvec^{(\rm pz)}$.
The displacement field is determined by the 
free-charge distribution  in the material: 
\be
    \nabla \cdot \Dvec  = e(p-n),
\label{divD}
\ee 
with $e$  the electron charge, and $p$ and $n$  the hole and
electron densities, respectively.
The evaluation of the electric field $\Evec$ inside
the semiconductor requires in general
 a self-consistent solution of Eq.~\ref{field} and
~\ref{divD} as outlined in Refs.~\onlinecite{ganapl}
and \onlinecite{noi.prb}.
However, many cases can be discussed without explicit calculations,
an useful exercise especially in the low free-carrier--density limit. 

An interesting such case is the evaluation of polarization 
effects in large, massive samples. In effect,
misunderstanding the role of the
 displacement field in massive samples leads to predicting
wrong  electrostatic fields in a MQWs system. 
At the outset, let it be pointed out 
that a strong polarization field in a III-V 
nitride massive sample does not imply the existence of 
a macroscopic electrostatic field over the whole sample. 
Qualitatively,  since such an electric field 
would produce (at non-zero  temperature) a persistent current due to
 thermally generated intrinsic carriers, and because this current cannot 
be sustained indefinitely, the electric field inside a massive sample
must be zero.  One  deduces thereforth that the displacement field in
the sample is  
\be
     \Dvec  = \Pvec^{(\rm sp)}.
\label{dvec}
\ee
For a neutral  (albeit polarized) sample,
 displacement  field conservation across the surface
would imply an electric field  $\Evec =\Pvec^{(\rm sp)}$  
in the vacuum off the sample surface. Of course this is unrealistic,
as  outside the sample $\Pvec$, $\Evec$, and $\Dvec$ must
all vanish. The SP discontinuity across the surface must be
 counterbalanced by an equal change in the displacement field.
Thus, near the surface of our sample, there exists a region
where  the free-carriers distribution satisfies
\be
    \int_{-\ell}^{0} 
\left(e(p-n) - \nabla \cdot \Pvec^{(\rm sp)}\right) dz= 0,
\ee
with $\ell$  the thickness of the surface region, which may range easily
in the hundreds of \AA.  
In this region the free carriers will constitute an approximately
 two-dimensional electron gas (2DEG) whose  areal 
density equals  the SP. 
The electric field in this region depends strongly on   material
parameters and on surface states. It can be concluded however that 
at distances in excess of $\ell$ 
from the surface (or, in the case of devices, the
interface with contacts, buffers, caps, etc.), the electric field will
be zero, and the polarization will equal
 $\Pvec^{(\rm sp)}$, as will the displacement field.  This 
is the starting point to discuss MQWs.
 
\section{MULTI-QUANTUM-WELLS}
As pointed out above (especially in a semi-insulating sample) 
 it can be safely assumed that the displacement field in a large sample
is uniform all over the crystal, and equal to $\Pvec^{(\rm sp)}$,
and $\Evec=0$. 
If a sequence of  QWs, of  fixed composition,
 is now inserted in the (otherwise  homogeneous)  sample,
the electric field inside the generic $j$-th layer (either  QW or
barrier) is given by:
\be
   \Evec_j = (\Dvec - \Pvec_j)/\eps_j
\label{qw}
\ee
with $\Pvec_j$  and $\eps_j$ the total polarization 
and dielectric constant in layer $j$.

If the displacement field is constant, {\it i.e.} if
 Eq. \ref{dvec} holds, the electric field in the $j$-th layer is
\be
\Evec_j = (\Pvec^{(\rm sp)} - \Pvec_j)/\eps_j,
\label{qw2}
\ee
where $\Pvec^{(\rm
sp)}$  is the SP in the massive sample material, which now functions
locally as barrier material. 
The field is then  zero in the barriers, which are made up of
unstrained material with spontaneous polarization 
$\Pvec^{(\rm sp)}$, while 
each  QW  in the series is subject
 to the same electric field 
$\Evec_j$. A potential drop given by 
\be 
\Delta V_j = |\Evec_j| \cdot l_j
\ee 
will thus occur in the generic $j$-th well of thickness $l_j$.
Now  Eq. \ref{dvec} holds only subject to the following
four  constraints:\\
{\it i)} the Debye-H\"uckel screening length is at least larger
than, say, the MQW region thickness;\\
{\it ii)} the QWs are far enough from any interface with the
outer world that the surface effects mentioned in 
the previous Section are negligible;   \\
{\it iii)} none of the $\Delta V_i$, nor their sum, must  exceed 
 the band gap energy of the well material:
\be\sum_k l_k \Evec_k < E_{\rm gap};
\label{sl2}
\ee
{\it iv)} the well-barrier interfaces are free of interface states. 

 These requirements are typically
met in nitride MQWs, except for condition {\it (iii)}, which breaks down
 if the QWs form a sufficiently thick superlattice.
Thereby, a self-consistent calculation explicitly
accounting for free carriers must be performed.
To approximately enforce condition $(iii)$ avoiding cumbersome 
calculations,
  it is convenient to 
approximate Eq. \ref{sl2} applying  periodic boundary conditions,
\be
    \sum_k l_k \Evec_k = 0,
\label{sl1}
\ee
where the sum runs over {\it all} layers in the MQW, 
 in particular  including the barrier layers.
Thereby the displacement field is determined subject to the condition
of zero average electric field in the MQW region. The maximal error in
the predicted  fields due to using Eq. \ref{sl1} instead of
Eq. \ref{sl2} is of the order of $E_{\rm gap}/d$, with $d$ the overall
thickness of the superlattice; hence this approximation becomes exact
in the limit of very thick superlattices.  The error may be of course
smaller depending on the specific conditions in the device (doping,
contacts, etc.).

Substituting Eq. \ref{qw}
 in Eq. \ref{sl1},
the displacement field is then obtained:
\be
    \Dvec = \frac{\sum_k l_k \Pvec_k/\eps_k}{\sum_k l_k/\eps_k}. 
\ee
Plugging this back into Eq. \ref{qw} gives
a simple expression for the field  in the generic well or barrier of the
MQW, namely
\be
\Evec_j =\frac{\sum_k l_k \Pvec_k/\eps_k - \Pvec_j \sum_k l_k/\eps_k}{\eps_j
\sum_k l_k/\eps_k},
\label{campo}
\ee 
with sums running on all layers (including the $j$-th).
This is a general expression for any thickness of wells and barriers
in a generic superlattice or MQW, which  reduces to the formulas in 
Ref. \cite{noi.prb} assuming a single type of well and barrier.

It is to be noted again  that in this case the fields are
non-zero {\it both} in the wells and the barriers. Typically,
the sign of the barrier field will be opposite to that of the well,
leading to an effectively triangular confinement potential,
which favors transfer of  confined electrons into the barrier layer
 on one side of the well, which may lead to anomalous 
effects in excitonic transitions  \cite{gil}.

In practice, if the conditions discussed above are satisfied
for a specific system,
the  properties of the latter can be calculated
by simply solving a Schr\"odinger equation for the
 compositional potential of the superlattice 
with the built-in fields as given by
Eq. \ref{campo}. Of course, this is emphatically not the case
in general, because of e.g. doping, excitation, boundary conditions, etc. 
A self-consistent solution
as outlined {\it e.g.} in Refs.~\onlinecite{ganapl}
and \onlinecite{noi.prb} will be needed.

\section{ISOLATED QUANTUM WELLS}
\label{sqw}
\begin{figure}[h]
\epsfclipon
\epsfxsize=14cm
\centerline{\epsffile{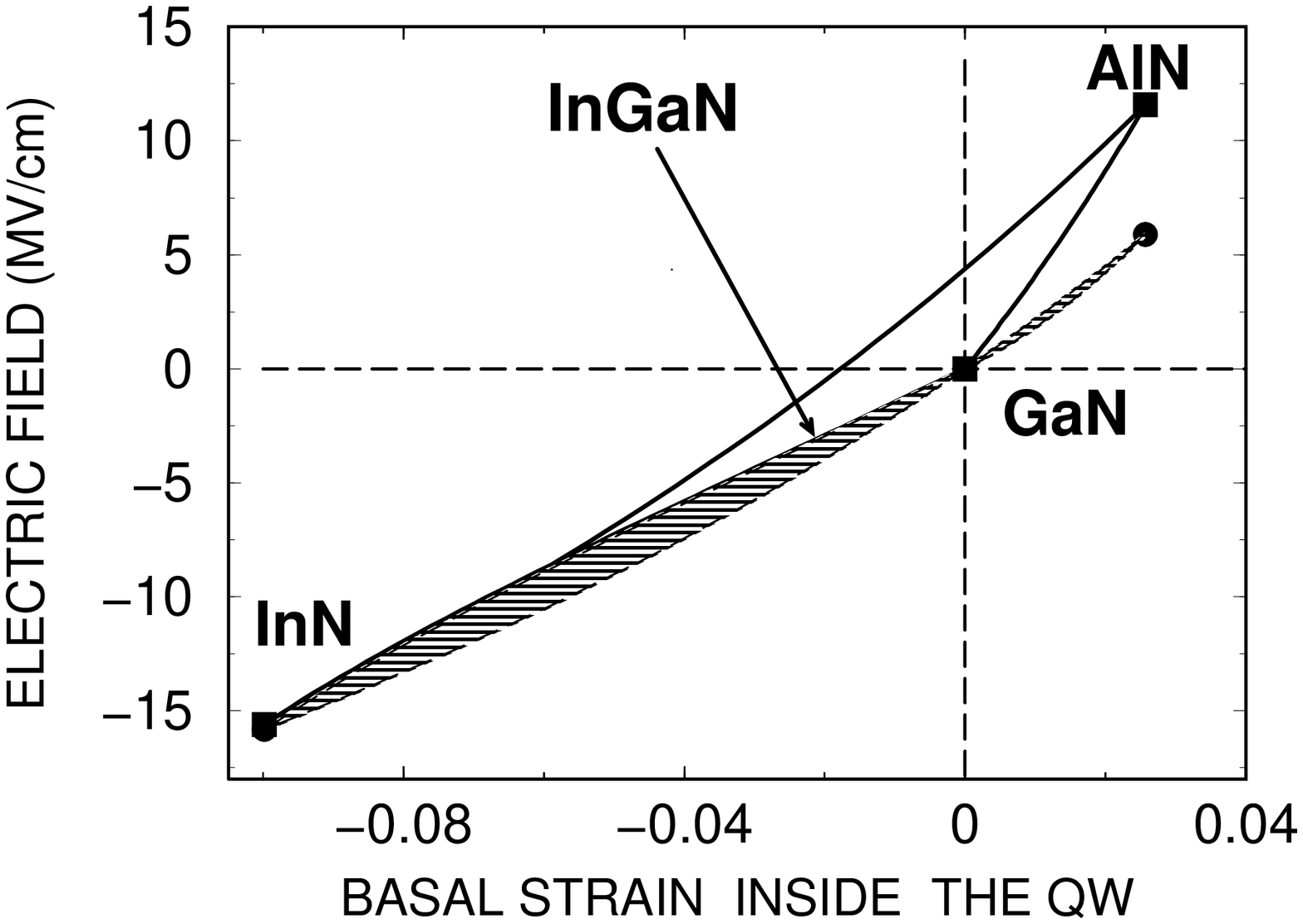}}
\vspace{3cm}
\caption{Electric field in an isolated
 Al$_x$In$_y$Ga$_{1-x-y}$N QW with GaN barriers as a function of the
well in-plane strain for all possible $x$ and $y$.
The white region delimited by the solid line 
encloses the accessible values of the  total electric field, the
 hatched region limited by dashed line encloses those of the purely
 piezoelectric polarization component. The two lines are barely
in contact at the upper edge of the hatched region: this contact zone, 
indicated by the arrow, corresponds to the InGaN alloy case (see text).}
\label{fig.sqw}
\end{figure}

Consider now a single  QW of material A;
 let the latter have dielectric constant $\eps_{\rm A}$ and
spontaneous polarization $\Pvec^{(\rm sp)}_{\rm A}$,
and the QW made thereof be
 strained so as to be subject to a piezoelectric polarization
$\Pvec^{(\rm pz)}$.
 The total polarization
in the QW is then  $\Pvec^{(\rm sp)}_{\rm A} + \Pvec^{(\rm pz)}$. 
The well is 
embedded in an extended, insulating,  and unstrained
material with spontaneous
polarization $\Pvec^{(\rm sp)}$. According to Eq.~\ref{qw},  
the field in the QW is 
\be
\Evec &=&  [\Pvec^{(\rm sp)} - \Pvec^{(\rm
sp)}_{\rm A}]/\eps_{\rm A} - \Pvec^{(\rm pz)}/\eps_{\rm A} \\\nonumber 
 &=& \Evec^{(\rm sp)}_{\rm A} + \Evec^{(\rm pz)}_{\rm A}.
\ee
At variance with conventional III-V's (having no SP),
an additional term $\Evec^{(\rm sp)}_{\rm A}$ appears,
due solely to the difference in the SP between 
the QW active layer and the barrier material  \cite{nota}. 
This additional term has far-reaching consequences, and 
can be used to tune the value of the electric field inside the QW.
In Fig.~\ref{fig.sqw} we depict
the  value of the electric field $\Evec$ and of its purely 
piezoelectric   component $\Evec^{(\rm pz)}$ vs in-plane strain 
for an AlInGaN QW embedded in GaN. Without SP, the values of the
electric field inside the QW  would fall into
the hatched  region delimited by dashed lines. The inclusion of 
SP gives rise to  a different, wider accessible region, especially for 
small  strains. The region in question is white and delimited by 
a continuous line. On the negative strain side of 
 Fig. \ref{fig.sqw}, this region barely touches
the upper side of the piezoelectric region, and shrinks becoming
a line for large strains. The various region boundaries are curved 
due to the quadratic dependence of the piezoelectric component on
alloy composition (through [piezoconstants]$\times$[strain] terms).

Several features are worth a mention. {\it First}, the two regions 
do not overlap, {\it i.e.} for any composition the 
total field differs from its pure piezoelectric component
(this of course is due to the fact that $\Evec^{(\rm sp)}$ never
vanishes). This difference is rather dramatic for positive strains, 
{\it i.e.} in Al-rich alloy wells. A  noticeable exception to 
this general behavior is that of
  GaN/InGaN QWs. These  correspond to the region where the total
and piezo field regions in Fig. \ref{fig.sqw} (almost) touch each
other:
in that case indeed, as InN has nearly the same SP as GaN, the SP difference
 between barrier and active layer, and hence $\Evec^{(\rm sp)}$,
is negligible with respect to the piezoelectric term.  
That is why early investigations on GaN/InGaN MQWs~\cite{takeuchi} 
  yielded results in good agreement with theory even though 
 SP was neglected. 
{\it Second}, without SP the  electric 
field is zero at zero strain (the typical situation in  zincblende
III-Vs).  This is no more the case in III-V nitrides, where
 SP is an additional degree of freedom potentially producing
 fields up to 4 MV/cm  {\it at zero strain}. This requires growing
AlInGaN with appropriate compositions, which appears to be
difficult because of thermodynamic solubility constraints; 
however, fields of several hundreds of KV/cm can be achieved
already at minute Al and In concentrations, which may perhaps
become accessible in the future.
{\it Third},  last but not least, SP provides a handle for reducing
or even, in principle, {\it  zeroing}
 the field  in strained QWs \cite{noi.prb}. Specifically,
this occurs on the zero field line in Fig. \ref{fig.sqw}. This can be a 
breakthrough for the many applications needing  a good confinement, but
at the same time no built-in  field in the active region. The same
considerations on growth hold as in the previous point.

\section{Summary and acknowledgements}
In conclusion,  SP can be considered as a degree of freedom 
to tune the value of the polarization-induced electric 
fields in QW systems.
It was shown, for a GaN/AlInGaN QW, that
fields up to 4MV/cm can be obtained 
in absence of strain, and that, conversely, a vanishing field can also be 
obtained, despite lattice mismatch, for judicious choices of
composition and strain.

Support from the PAISS program of INFM is acknowledged.  VF thanks
the Alexander von Humboldt-Stiftung for supporting his  stay at the
Walter Schottky Institut.

\end{multicols}

\begin{thebibliography}{99}
\bibitem{nota}
Note in passing that an equivalent result is obtained from
Eq. \ref{campo} in the limit of infinite barrier.
\end{thebibliography}
\end{document}